\documentclass[preprint,secnumarabic,amssymb, nobibnotes, aps, pra, showpacs,8pts]{revtex4-1}

\usepackage{graphics}
\usepackage{color}
\usepackage{subfig}
\usepackage{amsfonts}
\usepackage{amsmath}

\begin{document}
\title{Direct imaging of molecular symmetry by coherent anti-Stokes Raman scattering}

\author{Carsten Cleff $^1$}
\author{Alicja Gasecka $^{2-3}$}
\author{Patrick Ferrand $^1$}
\author{Herv\'e Rigneault $^1$}
\author{Sophie Brasselet $^1$}
\author{Julien Duboisset $^1$}
\email{julien.duboisset@fresnel.fr}

\affiliation{$^1$ Aix-Marseille Universit\'{e}, CNRS, Centrale Marseille, Institut Fresnel, UMR 7249, Domaine Universitaire de Saint J\'{e}r\^{o}me, F-13397 Marseille Cedex 20, France}
\affiliation{$^2$ Quebec Mental Health Institute Research Center, Laval University,
Qu\'{e}bec, Qc, G1J 2G3, Canada}
\affiliation{$^3$ Centre d'Optique, Photonique et Laser (COPL), Laval University,
Qu\'{e}bec, Qc, G1V 0A6, Canada}

\begin{abstract}
Nonlinear optical methods, such as coherent anti-Stokes Raman scattering (CARS) and stimulated Raman scattering (SRS), are able to perform label free imaging, with chemical bonds specificity. Here, we demonstrate that the use of circularly polarized light allows to retrieve not only the chemical nature but also the symmetry of the probed sample, in a single shot measurement. Our symmetry-resolved scheme offers simple access to the local organization of vibrational bonds and as a result provides enhanced image contrast for anisotropic samples as well as an improved chemical selectivity. We quantify the local organization of vibrational bonds on crystalline and biological samples, thus providing new information not accessible by spontaneous Raman and SRS techniques. This work stands for a novel symmetry-resolved contrast in vibrational microscopy, with potential application in biological diagnostic.
\end{abstract}

\maketitle

Obtaining information on the organization of matter on the micrometer-scale with non-destructive methods still remains a challenge in chemical physics and biology. One well-established method for extracting matter organization information is fluorescence microscopy, which uses fluorescent molecules or proteins to tag the sample. However, this technique is limited to the observation of the probe itself, which may differ from the sample organization.
Coherent Raman scattering (CRS) microscopy technique has proven to be powerful due to its label-free, three-dimensional, chemical selective and real-time imaging capabilities~\cite{Zumbusch1999, Potma2003, Min2011, Ji2014, Chen2015, Jung2015}. In coherent anti-stokes Raman scattering (CARS), two beams of different frequencies interact with the sample to excite a vibrational resonance. A probe beam is used to probe the vibrational excitation by generating a new anti-Stokes frequency shifted beam. Unfortunately, CARS is limited by a nonresonant four wave mixing (FWM) background resulting in reduced image contrast and chemical selectivity. Recently, it has been shown that stimulated Raman scattering (SRS) can provide vibrational spectra without nonresonant background allowing to report vibrational information with high fidelity and high efficiency ~\cite{Freudiger2008, Nandakumar2009, Ozeki2012, Berto2014, Camp2014}.
It is well known, that the symmetry properties of matter has a strong influence on its physical properties, e.g., in crystalline samples. Similarly, it has been found, that in biological environments anisotropic and symmetry properties of tissue are often related to specialized biological functionality. Nevertheless, the organization of molecular bonds in the focal volume, which is of particular interest in numerous situations where the medium is organized, is not contained in the spectral information. 
However, seminal works from nonlinear optics pioneers have shown that the Cartesian components of the nonlinear susceptibility tensor $\bar{\chi}^{(3)}$ express the vibrational symmetry properties~\cite{Maker1970, Jerphagnon1978}. To read the tensor elements, polarization resolved schemes have been proposed decades ago~\cite{Oudar1979}, stimulating more recent developments in microscopy~\cite{Cheng2001, Lu2008, Munhoz2010, Zimmerley2013, Upputuri2013, Bioud2014}. 
The molecular organization from a sample is usually retrieved by acquiring a stack of images from different polarization angles of the excitation or detection light fields, requiring long acquisition times, time consuming post-processing and introducing some artifacts ~\cite{Duboisset2014a}. 

In this article, we introduce a new label-free microscopy technique that is able to retrieve the individual symmetry orders of molecular organization in a single shot measurement. The symmetry-resolved CARS (SR-CARS) signal not only depends on the presence of molecular bonds, but also on their organization within the focal volume. By switching between combination of left- and right-handed circular polarization states for the involved fields, it is possible to directly image individual symmetry contributions of the sample. 
Beyond this new chemical selectivity, we show that our technique can (1) suppress the background in CARS images and spectra, thus enhancing the contrast by one to two orders of magnitude, and (2) retrieve quantitative information without pre-knowledge information on the molecular organization, without post-processing and independently of sample orientation in the transverse plane. SR-CARS provides higher chemical selectivity based on different symmetry characteristics, which are not accessible with regular spontaneous Raman or SRS microscopy.\\

%++++++++++++++++++++++++++++++++++++++++++++++++++++++++++++++++++
\textbf{\textcolor{black}{{Light matter symmetry matching}}}
%++++++++++++++++++++++++++++++++++++++++++++++++++++++++++++++++++
A direct read out of a specific sample symmetry is possible when the light field tensor probing the sample only consists of the targeted symmetry of the sample~\cite{Yuratich1977, Brasselet1998}. In practice, the detected CARS electric field amplitude $E_{as}$ (anti-Stokes) along a specific polarization direction is the projection of a tensor $\bar{F}$, representing all involved light fields, onto the $\bar{\chi}^{(3)}$ susceptibility tensor of the medium
\begin{equation}
E_{as} = \bar{\chi}^{(3)} \cdot \bar{F}  =
\bar{\chi}^{(3)} \cdot (\hat{e^*}_{as} \otimes \vec{E}_{p} \otimes \vec{E}^*_{s} \otimes \vec{E}_{pr} ), 
\label{eq:cars_cartesian}
\end{equation}
where $\otimes$ is the dyadic product and $^*$ stands for the complex conjugate, $\vec{E}_{p}$, $\vec{E}_{pr}$  and $\vec{E}_{s}$ are the pump, probe and Stokes fields, respectively, and $\hat{e}_{as}$ is the unit vector along the polarization direction of the emitted anti-Stokes field, such that the four vector fields create a rank-four tensor~\cite{Duboisset2014}. 

As our technique targets the investigation of sample's symmetry, it is convenient to choose spherical coordinates ($\theta$, $\phi$), for which left- ($E_{\circlearrowleft}$) and right-handed ($E_{\circlearrowright}$) circular polarization and linear $z$-polarization states ($E_{z\rightarrow}$) can be described by a set of spherical harmonic functions $Y^{l}_{m}(\theta,\phi)$ with $l = 1$
\begin{eqnarray*}
E_{\circlearrowleft}(\theta,\phi) = aE_{0,\circlearrowleft} \cdot Y^{1}_{1}(\theta,\phi) \\
E_{\circlearrowright}(\theta,\phi) = aE_{0,\circlearrowright} \cdot Y^{1}_{-1}(\theta,\phi) \\
E_{z\rightarrow}(\theta,\phi) = aE_{0,z\rightarrow} \cdot Y^{1}_{0}(\theta,\phi)
\label{eq:polar_circular}
\end{eqnarray*}
where $E_{0,\circlearrowleft}$, $E_{0,\circlearrowright}$, $E_{0,\rightarrow}$ are the electric field amplitudes associated to the quantum number $m=1$, $m=-1$ and $m=0$, respectively and $a=-\sqrt{{4\pi} /{3}}$ is a normalization factor (Supplementary Section 1).

The advantage of using spherical harmonic functions is that an $m$-value is a direct reporter of the presence of an $m$-folded rotational invariant symmetry in the sample plane. In case of the CARS field tensor $\bar{F}$ in equation (\ref{eq:cars_cartesian}), the multiplication of spherical harmonic functions generates a new set of spherical harmonic functions of fixed $m$-values, allowing to read $m$-values higher than the initial individual light fields and to determine the symmetry of the combined light fields
\begin{equation*}
\bar{F}(\theta,\phi) \propto Y^{1}_{m_p} \cdot Y^{1^{*}}_{m_{s}} \cdot Y^{1}_{m_{pr}} \cdot Y^{1^{*}}_{m_{as}} =  \sum_{l = 0}^{4} k_l \cdot Y^{l}_{m_{\bar{F}}},
\label{eq:CARS_spherical}
\end{equation*}
with the resulting $m_{\bar{F}}$-value being a summation of field's $m$-values
\begin{equation}
m_{\bar{F}} = m_p - m_{s} + m_{pr} - m_{as},
\label{eq:mvalue}
\end{equation}
where $k_l$ are the Clebsch-Gordan coefficients weighting the different spherical harmonic functions (Supplementary Section 1) and $Y^{l*}_{m} = Y^{l}_{-m}$. Therefore, the light field tensor $\bar{F}(\theta,\phi)$ has a $m_{\bar{F}}$-rotational symmetry in the sample plane defined by the $m$-values of the individual incident and emitted light fields, see equation~(\ref{eq:mvalue}).

When $\bar{F}$ (light) probes $\bar{\chi}^{(3)}$ (matter) in a CARS process (equation~(\ref{eq:cars_cartesian})), it can be seen, from the orthogonality of spherical harmonic functions, that $\bar{F}$ only probes parts of the $\bar{\chi}^{(3)}$-tensor with identical rotational invariant symmetries (i.e. identical  $m_{\bar{F}}$). Thus, engineering of $\bar{F}$ allows the direct read-out of specific sample symmetries, creating a symmetry based contrast mechanism.

\begin{figure*}[t!]
%\begin{center}
	\includegraphics{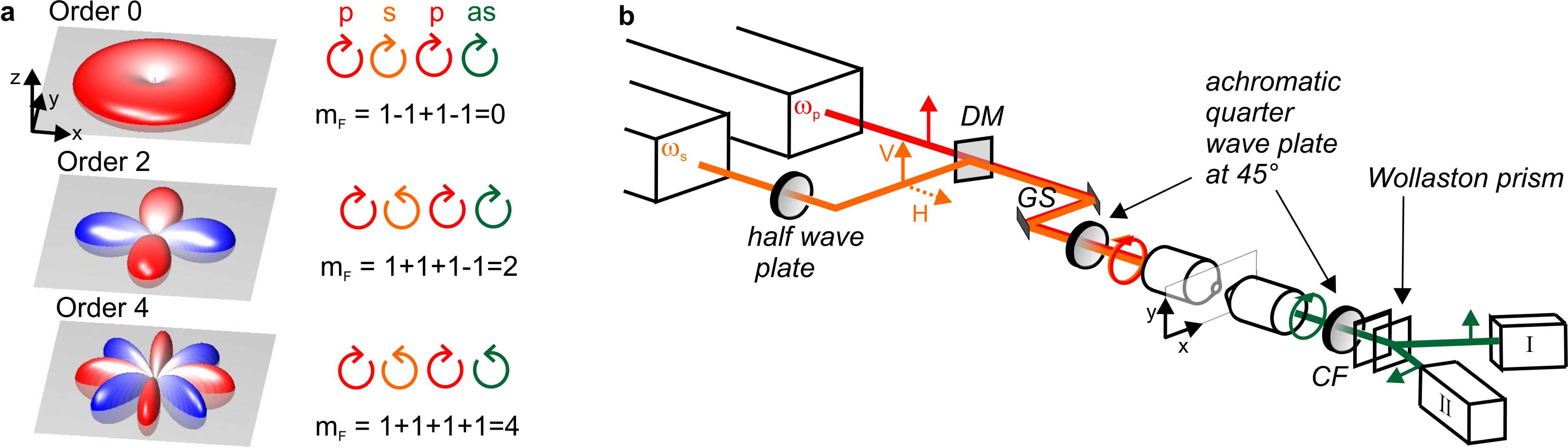}
	\caption{SR-CARS. (a) Generated CARS light field $\bar{F}(\theta,\phi)$ with symmetry orders $m_{\bar{F}} = m_p - m_{s} + m_{p} - m_{as}$ resulting from the combination of circular polarization states for the pump (red), Stokes (orange) and detected CARS (green) fields; (b) schematic of the experimental setup (GS: galvo-scanners; DM: dichroic mirror; CF: CARS filter), switching the Stokes field between $V$- and $H$-polarization states before the first quarter wave plate allows switching between Stokes field $\circlearrowleft$- and $\circlearrowright$-polarization at the sample plane. Using two photomultiplier tubes (I/II) detecting $\circlearrowleft$- and $\circlearrowright$-polarized CARS fields allows to target the different symmetry orders  $m_{\bar{F}}$=0, 2 and 4.\label{fig1}}
%\end{center}
\end{figure*}

Figure~\ref{fig1}-a displays the different light fields $\bar{F}(\theta,\phi)$ of symmetry orders $m_{\bar{F}}$ that can be generated using circularly polarized light and degenerated CARS ($m_p = m_{pr}$). As circular polarizations provide only $m=1$ or $m=-1$ values, sample rotational symmetry contribution of order 0 ($m_{\bar{F}}=0$), order 2 ($m_{\bar{F}}=2$), and order 4 ($m_{\bar{F}}=4$) can be specifically addressed. Using two detectors for the emitted $\circlearrowright$ and $\circlearrowleft$ CARS polarization states allows the independent and simultaneous detection of symmetry contributions of order 0 (isotropic) and 2 (two-fold symmetry) respectively, if the pump, Stokes, and probe beams have the same circular polarization states (Fig. \ref{fig1}-b). If the Stokes circular polarization state is changed from $\circlearrowright$ to $\circlearrowleft$, symmetries of order 2 and 4 (four-fold) are then probed independently and simultaneously. Using circularly polarized light at both excitation and detection paths in CARS then creates a new symmetry resolved imaging modality that has the unique ability to extract local orientational symmetry information from any sample. The orders $m_{\bar{F}}$=1 and $m_{\bar{F}}$=3 are in principle also accessible, but require excitation or detection of $z$-polarization contributions, which is not considered in this work. Note that since the obtained information is invariant upon rotation around the $z$ axis, each symmetry order image does not depend on the sample orientation in the sample plane.\\

%++++++++++++++++++++++++++++++++++++++++++++++++++++++++++++++++++
\textbf{\textcolor{black}{{Symmetry resolved imaging}}}

\begin{figure*}[t!]
%\begin{center}
	\includegraphics{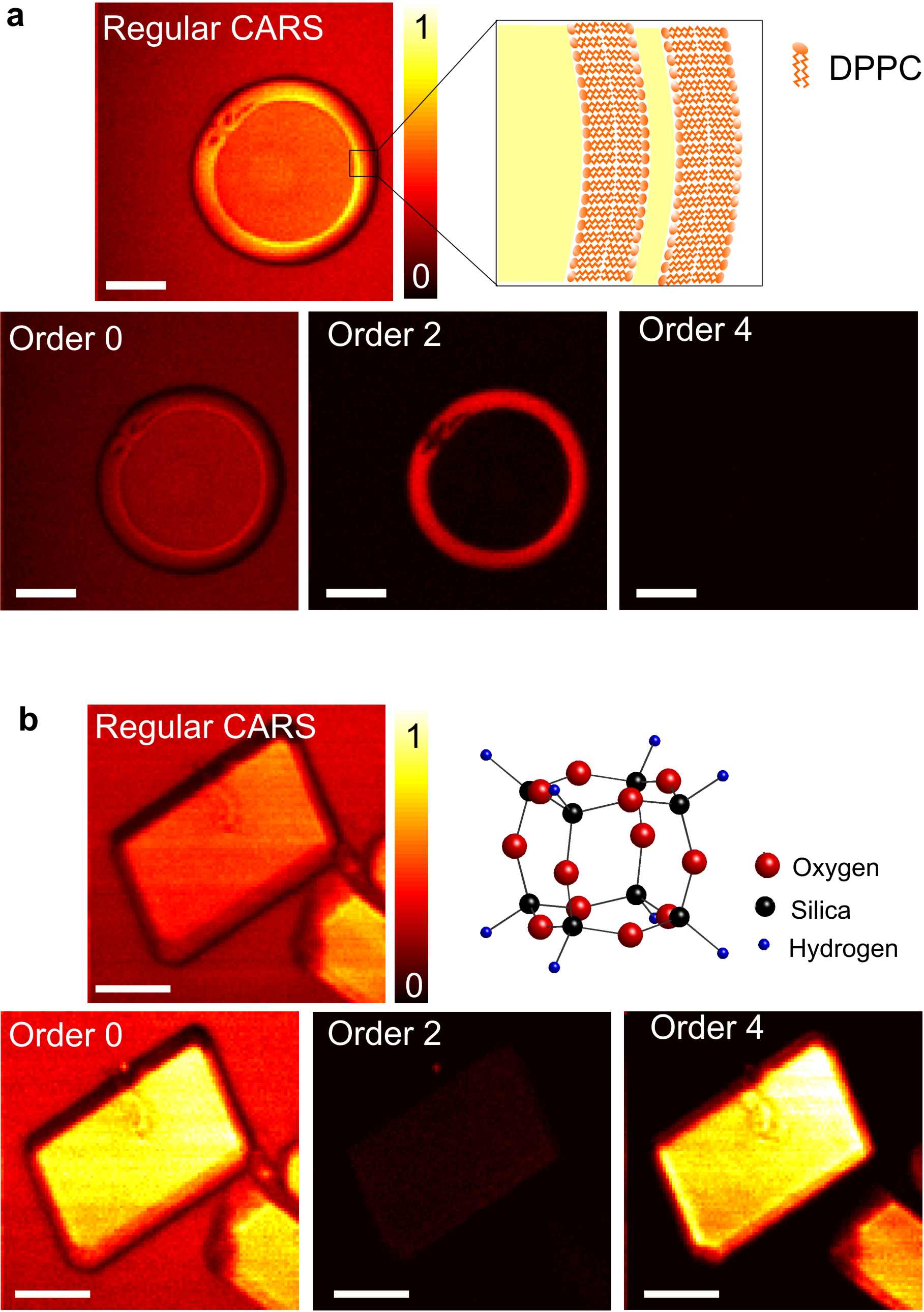}
	\caption{SR-CARS imaging. Regular and SR-CARS images of (a) a multi lamellar DPPC vesicle (MLV) at 1133~$\mathrm{cm}^{-1}$ and (b) a HT8 zeolite crystal at 932~$\mathrm{cm}^{-1}$. Schematic representation of the membrane structure of a MLV and the cubic unit cell of HT8 zeolite are also shown. Scale bar $10\mu m$\label{fig1_2} }
%\end{center}
\end{figure*}

\textbf{\textcolor{black}{{Multi lamellar vesicles.}}}
%++++++++++++++++++++++++++++++++++++++++++++++++++++++++++++++++++
To demonstrate the potential of SR-CARS imaging, we applied our scheme to model samples of known symmetry. Dipalmitoylphosphatidylcholine (DPPC) multi lamellar lipid vesicles (MLVs) are made of a tight packing of lipid layers, which form a ring of highly order matter with two-folded symmetry (see drawing in Figure~\ref{fig1_2})  and a lipid orientational distribution close to a Gaussian angular shape~\cite{Bioud2014}. Regular CARS imaging of MLVs on C-H stretching vibration at 1133~$\mathrm{cm}^{-1}$ embedded in water, shows a poorly contrasted image due to the nonresonant background, see Figure~\ref{fig1_2}-a. Using SR-CARS imaging, different symmetry contribution can be separately imaged, namely the isotropic ($m_{\bar{F}}$=0), two-fold ($m_{\bar{F}}$=2) and four-fold ($m_{\bar{F}}$=4) contributions of the equatorial section of the MLV in the sample plane. Figure~\ref{fig1_2}-a shows that the aqueous solution surrounding the MLV is only visible in the order 0 image, due to its purely isotropic nature. The microscopic organization of lipids results in a strong order 2 signal, allowing  background-free imaging of the MLV with a clearly superior contrast respect to regular CARS image. At last, no visible order 4 is present, meaning that the lipid orientational distribution is rather smooth, such as in a Gaussian shape, in contrast to sharp distributions that would involve higher symmetry harmonics~\cite{Bioud2014}. To obtain more quantitative information about the molecular organization, the order 2 image is normalized by the total intensity image (sum over all the order images) pixel by pixel and square rooted, leading to a direct read-out of the order 2 susceptibility tensor contribution. This normalized order 2 is found to be around 0.5 (Supplementary Section 3) and is in good agreement with previous results obtained in similar systems~\cite{Bioud2014}.

\textbf{\textcolor{black}{{Zeolite crystal.}}}
 Since molecular vibrational resonances belong to a variety of symmetry groups, SR-CARS can be used to retrieve the vibration symmetries. As a second model system we selected the cubic crystal octahydrosilasesquioxane $H_{8}Si_{8}O_{12}$ crystal (HT8) zeolite, which forms microscopic scale crystals belonging to the $O_h$ crystalline point group, and has a four-fold symmetry for the 932~$\mathrm{cm}^{-1}$ O-Si-H vibrational resonance addressed here ~\cite{Marcolli1997}. Regular CARS imaging of such a crystal embedded in water shows a poorly contrasted image (Fig.~\ref{fig1_2}-b) due to the nonresonant background. Employing SR-CARS, the surrounding background disappears in the order 4 image, enhancing the CARS image contrast of the crystal with respect to its isotropic surrounding by a factor of 100. The order 2 image, as expected from a pure four-fold symmetry, shows no visible signal.
 
\begin{figure*}[t!]
\begin{center}
	\includegraphics{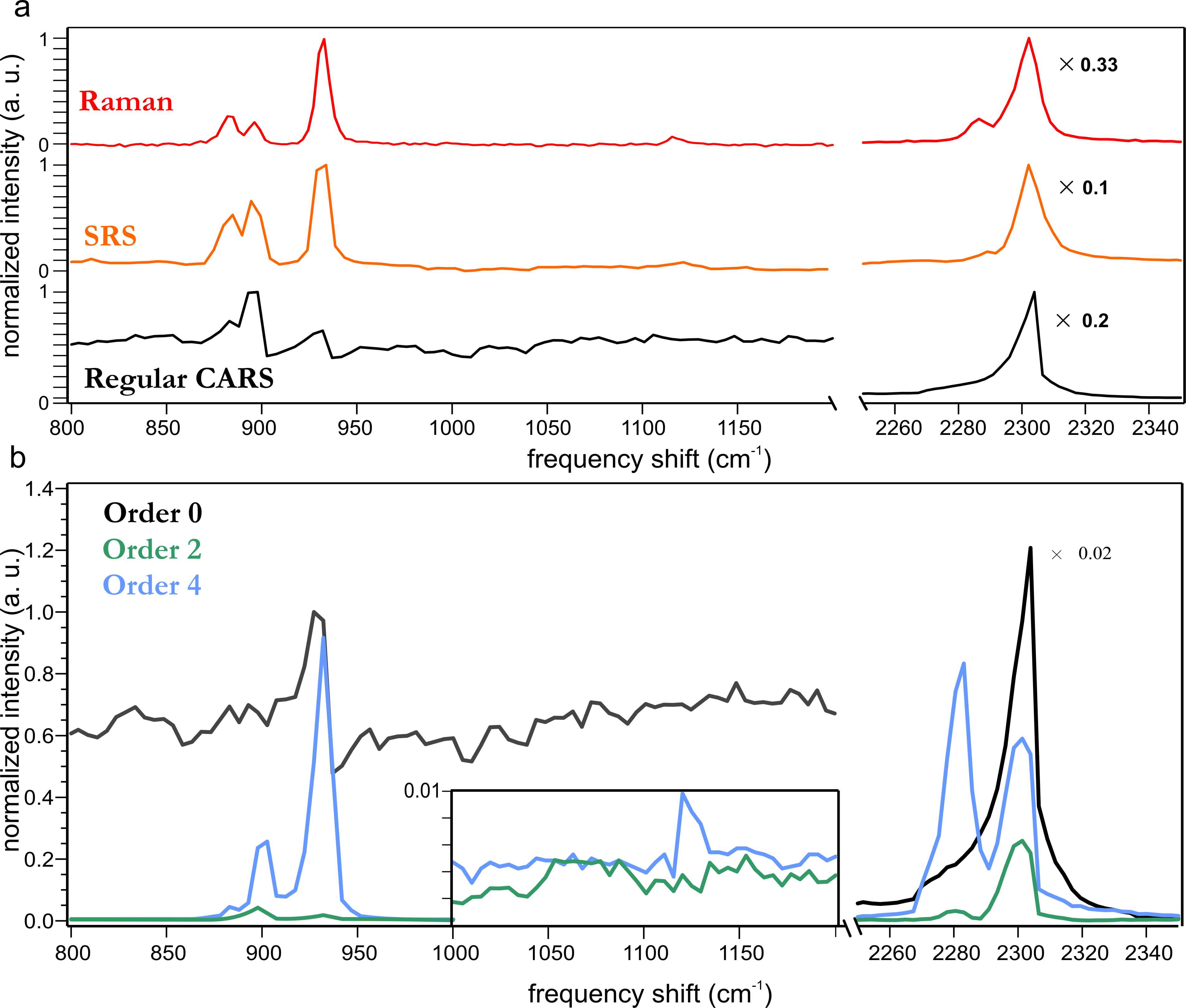}
	\caption{SR-CARS spectrum. Vibrational spectra of a HT8 zeolite crystal obtained with spontaneous Raman, SRS, regular CARS and SR-CARS order 0, 2 and 4. \label{fig:Zeolite:SpectrumAndImage}}
\end{center}
\end{figure*}

SR-CARS can be used to increase spectral contrast and separate close-by or overlapping resonances. Figure~\ref{fig:Zeolite:SpectrumAndImage}-a shows the vibrational spectrum (obtained by scanning the Stokes beam wavelength) of a HT8 zeolite crystal acquired with spontaneous Raman scattering, SRS and regular CARS. In comparison to spontaneous Raman and SRS spectra, regular CARS is clearly affected by a nonresonant background generated by the crystal itself, which prevents from visualizing all the resonances of HT8 in the scanned spectral range. 

Figure~\ref{fig:Zeolite:SpectrumAndImage}-b displays order 0, order 2 and order 4 spectra acquired with the SR-CARS scheme. The first observation is that order 4 provides a highly contrasted background-free spectrum. It reveals in particular a weak vibrational resonance at 1120~$\mathrm{cm}^{-1}$ that is invisible in regular CARS. Moreover, it clearly separates resonances that are less distinct in Raman or SRS (around 2300~$\mathrm{cm}^{-1}$). The separation of overlapping vibrational resonances with SR-CARS is possible, as the resonances are of different vibrational mode types with different symmetry properties. As a result, they have similar amplitudes for their four-folded symmetry contributions (thus allowing to separate them) while the total amplitudes of the two resonances are of different orders of magnitude such that they can hardly be distinguished in spontaneous Raman or SRS. While regular CARS of order 0 spectra are clearly distorted by the intrinsic interference with the nonresonant background (which is itself primarily of order 0), the high order symmetry filtering extracts much clearer spectral information (see also the whole spectral scan in Supplementary Movie M1). The order 2 spectrum shows weak signals, which we attribute to residual polarization leakage. 
The second observation is that all the resonances of the crystal don't have the same symmetry-resolved spectra. The $A_{1g}$ resonance at $2302~\mathrm{cm}^{-1}$ has mainly order 0 contribution,  the order four is less than 1\% of the isotropic order and come from optical leakage. The $T_{2g}$ resonances at $883~\mathrm{cm}^{-1}, 897~\mathrm{cm}^{-1}, 1117~\mathrm{cm}^{-1}, 2286~\mathrm{cm}^{-1}$  have only a strong order 4 contribution and the $E_{g}$ resonance at $932~\mathrm{cm}^{-1}$ has both order 0 and order 4 contributions. The decomposition of each mode on the symmetry orders is determined by the selection rules and detailed in the Supplementary information S2. The three different modes of zeolite investigated here have three different signatures on the SR-CARS spectra, allowing an unambiguous experimental identification of each mode. 

\textbf{\textcolor{black}{{Myelin sheaths.}}}
We now focus on myelin sheaths, that are highly organized multilayer membranes made of lipids and proteins surrounding the axon of neurons. Myelin integrity is essential for the propagation of action potential and CARS has proven to be a powerful label free technique for myelin imaging, targeting lipid vibrational bonds~\cite{Wang2005,Belanger2009}. Polarization resolved CARS has been reported to describe myelin molecular organization~\cite{Bioud2014}, and to relate this organization to demyelination processes~\cite{deVito2014}, however such experiments take minutes as they require multiple images with different incoming linear polarization states.  Here, we apply SR-CARS imaging to mice spinal cords and show that it can enhance information and reveal the full molecular lipid organization in single shot images, within biological tissues.

\begin{figure*}[t!]
\begin{center}
	\includegraphics{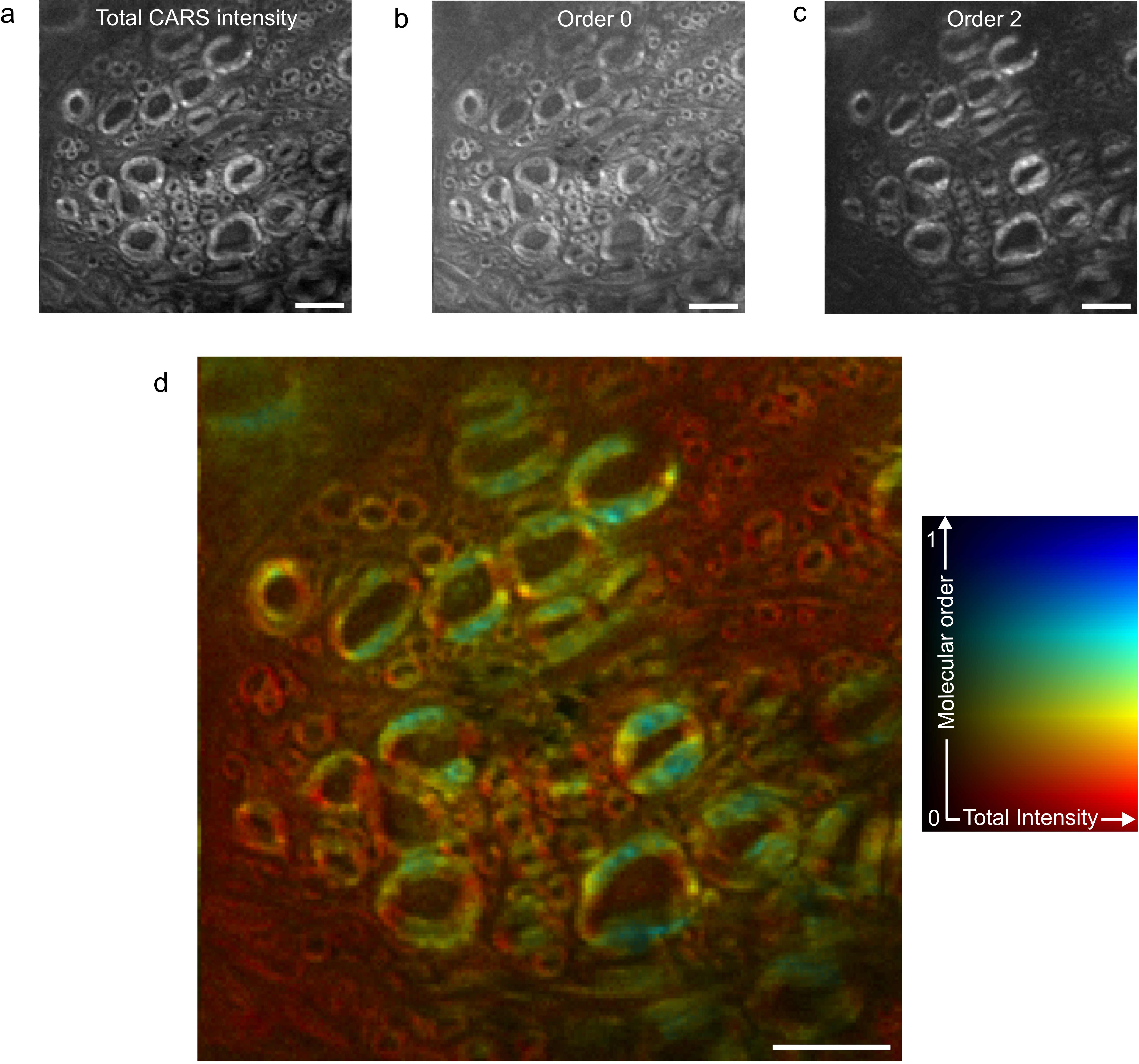}
	\caption{SR-CARS imaging in a biological tissue (\emph{ex-vivo} myelin sheaths in mice spinal cord) targeting the Carbon-Carbon vibrational stretching mode of lipid chains at 1099~$\mathrm{cm}^{-1}$. (a) Total CARS intensity (circular excitation and unpolarized detection). (b) SR-CARS image of order 0, exhibiting isotropic symmetry contributions in the sample. (c) SR-CARS image of order 2, highlighting the two-fold symmetry CARS contribution associated to organized lipids. (d) Quantification of Local molecular order in the myelin sheath using normalized order 2 image with respect to the total intensity image, pixel by pixel. The local molecular order is color coded, from 0 (for completely disordered) to 1 (for fully oriented), while the total intensity is encoded in the pixel brightness. scale bar 5$\mu m$. \label{fig:ColorCoded}}
\end{center}
\end{figure*}

Figure~\ref{fig:ColorCoded}-a shows a CARS intensity image of an \emph{ex-vivo} mouse spinal cord transverse section, using circular excitation on the C-C vibrational bond at 1099~$\mathrm{cm}^{-1}$. The vibration is chosen for its weak efficiency, as compared to CH2 stretching modes, to demonstrate the performance of SR-CARS scheme. The order 0 image (Fig.~\ref{fig:ColorCoded}-b), shows weak contrast due to the presence of lipids bonds surrounding the myelin sheath and an isotropic contribution of the nonresonant background. On the contrary, the order 2 image (Fig.\ref{fig:ColorCoded}-c) highlights the ring shaped myelin around the axons, exhibiting the fact that lipids are highly organized and oriented in such structures. The order 4 image shows very weak signals (not shown), similarly as in MLVs. To provide a quantitative information of molecular order in myelin sheath using SR-CARS, the order 2 image is normalized by the total intensity image (sum of all orders) pixel by pixel, in order to cancel the molecular density dependency, see Fig.~\ref{fig:ColorCoded}-d. The molecular order of lipids in myelin sheaths is visibly not homogeneously distributed, with highly organized regions like in MLV can be found (high order 2 values) as well as highly disordered regions with low order 2  (Fig.~\ref{fig1_2}-a). This heterogeneity can be attributed to different morphologies and molecular compositions present in multilayers surrounding axons, which is a topic of current interest to address possible molecular-level neurobiology diagnostics~\cite{deVito2014}. The symmetry resolved modality developed here is then able, in a single image allowing for high speed imaging, to reveal local heterogeneities that are absent from the regular CARS images.\\

%++++++++++++++++++++++++++++++++++++++++++++++++++++++++++++++++++
\textbf{\textcolor{black}{{Discussion}}}
%++++++++++++++++++++++++++++++++++++++++++++++++++++++++++++++++++

This work presents a new CARS imaging modality based on symmetry-selectivity, with unprecedented ability to enhance vibrational contrast and to reveal molecular bonds organization symmetry. The use of circular polarization makes this imaging modality independent of the sample orientation in the transverse sample plane, making this contrast enhancement efficient without the need to find an optimal polarization coupling direction.

We exploited the specific symmetry of molecular bonds assemblies to strongly enhance the CARS image contrast in crystalline and tissue sample. In the past, many other techniques have been presented to improve contrast in CARS, though they mostly aimed to reduce the nonresonant background. A symmetry-based approach has the strong advantage of improving  image contrast independently of the resonant or nonresonant nature of the background. 

Finally, molecular orientational organization imaging can be exploited to achieve a novel type of structural contrast in biological samples, which can improve monitoring of biological processes and diseases as well as diagnostics. This provides, for the first time, quantitative molecular bond symmetry imaging in a single image acquisition without the need for any polarizer rotation nor signal processing. Biochemical studies have shown that molecular organization in myelin sheaths is indeed highly affected in diseases like multiple sclerosis and leukodystrophies, however no current imaging technique can quantitatively address this issue \emph{in vivo}. Investigating biological functions related to molecular organization in real time would be highly valuable and SR-CARS could be the basis of high-speed diagnostics. \\

 %++++++++++++++++++++++++++++++++++++++++++++++++++++++++++++++++++
\textbf{\textcolor{black}{{Methods}}}
%++++++++++++++++++++++++++++++++++++++++++++++++++++++++++++++++++

\textbf{CARS microscope.}
CARS imaging was performed on a custom-built setup incorporating a picosecond stimulated Raman optical source~\cite{Brustlein2011}. This source is composed of two optical parametric oscillators (OPO1 and OPO2, Emerald, APE) synchronously pumped by a mode-lock frequency doubled Nd:YVO Laser (PicoTrain, HighQLaser) operating at 532nm. The two mode locked beams from OPO1 (pump) and OPO2 (Stokes) (pulse duration 5 ps, repetition rate 76 MHz) are overlapped in time and space and sent into a custom made scanning microscope. SR-CARS signals (order 0, 2 and 4) are detected in the forward direction using PMTs (Hamamatsu, H10682) working in photon-counting regime.  Excitation and collection are provided by an NA=0.6 objective (Olympus UCPlan FL 40x). Incident powers at the sample plane were 1-4 mW for the pump beam and 1-5 mW for the Stokes beam depending on the samples. Imaging is performed by scanning galvanometric
mirrors (typically: pixel dwell time of 50 $\mu s$, 100 $\times$ 100 pixels, and scan range of 30 $\mu m$). A dedicated software controls the galvo mirrors, the acquisition card, and the OPO wavelength tuning for spectral acquisition~\cite{Ferrand2015}.

In order to achieve SR-CARS imaging, an achromatic quarter-wave plate was inserted before the focusing microscope objective to excite the sample with circularly polarized light. The linear polarization state (before the quarter-wave plate) of the Stokes beam was switched between V- and H-polarization resulting in a switching between $\circlearrowleft$ and $\circlearrowright$-circular polarization states in the sample plane. The generated CARS signal passed through a second achromatic quarter-wave plate converting circular polarization to linear polarization. Subsequently, a Wollaston prism split the CARS beam into V- and H-polarization, which were detected individually with photomultiplier tubes. Consequently, the two photomultiplier tubes were sensitive to $\circlearrowleft$ and $\circlearrowright$-circular polarization states in the sample plane.

In the SRS operation mode, the pump beam was modulated in amplitude at a frequency of 20 MHz by an acousto-optic modulator (AOM - AAOptoelectronic MT200 A0,2-800). The stimulated Raman gain (SRG) induced on the Stokes beam was detected in the forward direction by means of a high speed photodiode and a fast lock-in amplifier (manufactured by APE) in a way similar to ~\cite{Freudiger2008}.

The Raman spectrum was acquired using a HeNe laser at 632.8 nm and a spectrograph (Horiba iHR320) equipped with a Peltier-cooled CCD detector.

\textbf{MLV samples.}
MLVs were made from chain 1,2-dipalmitoyl-sn-glycero-3-phosphocholine (DPPC) lipids and 5\% cholesterol. DPPC was
hydrated in phosphate-buffered saline [(PBS), pH 7.4]  above the main phase-transition temperature (45$^{\circ}$C) for 1 h, leading
to MLVs of 1-30$\mu m$ size in a solution enclosed between two spaced coverslips. The MLVs form almost spherical objects made of concentric multilayers of lipids. The images were performed at the equatorial plane of these objects where the distribution of the lipids is expected to be lying along the transverse sample plane.

\textbf{Zeolite samples.}
The sample studied in this work is an octahydrosilasesquioxane $H_{8}Si_{8}O_{12}$ crystal (HT8), that has cubic symmetry
and belongs to the O$_{h}$ crystallographic point group. The $H_{8}Si_{8}O_{12}$ crystal synthesis can be found in~\cite{Marcolli1997}. For CARS imaging, micrometric to millimetric size crystals are directly deposited onto microscope coverslip and surrounded by water.

\textbf{Myelin samples.}
Mice were anesthetized and perfused intracardially with 4\% paraformaldehyde (PFA, Fischer Scientific, Pittsburgh) in 0.1 M phosphate bufer (PB). The whole spinal cord was dissected out from each mouse and placed in 4\% PFA overnight. Lumbar spinal segments corresponding to L3-5 level were isolated from the spinal cord, rinsed several times with 0.1 M PB. 300 $\mu m$-thick slices were cut in the sagittal plane using a vibratome (Leica, VT 1000).

\textbf{\textcolor{black}{{Acknowledgements}}}
We thank G. Calzaferri (University of Bern) and P. A. Agaskar (Spherosils LLC, Blacksburg, VA) for providing us with the HT8 crystals. We thank D. Cot\'e (University of Laval) for providing us with the mice myelin sheats slices.
The authors acknowledge financial support from the Centre National de la Recherche Scientifique (CNRS), Aix-Marseille University A*Midex (noANR-11-IDEX- 0001-02), ANR grants France Bio Imaging (ANR-10-INSB-04-01) and France Life Imaging (ANR-11-INSB-0006) infrastructure networks and 'Fondation pour la Recherche M\'edicale' grant DBS20131128448.

\textbf{\textcolor{black}{{Author contributions}}}
C.C. performed all experiments. A.G. prepared the sample, S.B., H.R and J.D. designed all experiments and developed the signal-processing methodology and protocols. C.C. and J.D analyzed the results. P.F. developed the acquisition and processing software. C.C, J.D., S.B. and H.R. wrote the manuscript.

\textbf{\textcolor{black}{{Additional information}}}
Movie M1 shows symmetry resolved spectroscopy and hyper-spectral imaging of HT8 zeolite crystal. For each wavenumbers, regular CARS, order 0 and order 4 are shown. Supplementary information is available in the online version of the paper.

\textbf{\textcolor{black}{{Competing financial interests}}}
The authors declare no competing financial interests.

\newpage

\newpage

\newpage

\newpage

\end{document}